\documentclass[preprint,showpacs,preprintnumbers,amsmath,amssymb]{revtex4}

\usepackage{graphicx}% Include figure files
\usepackage{dcolumn}% Align table columns on decimal point
\usepackage{bm}% bold math
\usepackage{booktabs}
\usepackage{amssymb,mathrsfs,bbm,amscd}
\usepackage{fancyhdr}
\usepackage{lastpage}
\begin{document}
\fancyhf{}
\fancyhead[c]{\small Submitted to Chinese Physics C}

\title{Whole fusion-fission process with Langevin approach and compared with analytical solution for
barrier passage}

\author{Jie Han and Jing-Dong Bao\footnote{Corresponding author: jdbao@bnu.edu.cn}}

\affiliation{Department of Physics,Beijing Normal University,
Beijing 100875, People's Republic of China}

\begin{abstract}
We investigate time-dependent probability for a Brownian particle
passing over the barrier to stay at a metastable potential pocket
against escaping over the barrier. This is related to whole
fusion-fission dynamical process and can be called the reverse
Kramers problem. By the passing probability over the saddle point of
inverse harmonic potential multiplying the exponential decay factor
of a particle in the metastable potential, we present an approximate
expression for the modified passing probability over the barrier, in
which the effect of reflection boundary of potential is taken into
account. Our analytical result and Langevin Monte-Carlo simulation
show that the probability passing and against escaping over the
barrier is a non-monotonous function of time and its maximal value
is less than the stationary result of passing probability over the
saddle point of inverse harmonic potential.
%We also calculate the
%stationary value of time-dependent passing probability as a function
%of different initial conditions.

\pacs{25.60.Pj, 24.60.Ky, 05.40.Jc}
\end{abstract}

\maketitle

\section{\label{sec:level1}introduction}

The  metastable system decay can be applied widely to describe
various science problems such as chemical reaction kinetics, phase
transient, nuclear fission, and so on. The well-known Kramers
problem is such a process that a Brownian particle subjected to
thermal fluctuation escapes from the barrier of a metastable
potential. As early as 1940, Kramers published his seminal paper
``Brownian motion in force fields and chemical reaction diffusion
model'' \cite{HA1940}, in which he proposed a formula for the
reaction rate constant for a general-damped particle escaping from
a metastable potential well and used this model to explain the
mechanism of excited nuclear fission. Abe is the first researcher
who used Langevin Monte-Carlo simulation to calculate numerically
the nuclear fission rate \cite{Abe1986}. In 1990, H\"anggi
\emph{et. al.} \cite{PH1990} summarized the works fifty years
after Kramers, including various improvements and extensions for
the Kramers rate theory.

Now a reverse problem appears timely, i.e., a Brownian particle with
initial velocity passes over the saddle point to enter into the well
of metastable potential and escapes from the saddle point finally.
In fact, molecular collision, atom cluster and heavy-ion fusion are
such barrier passage problems \cite{YA1997, YY1997, LB2006, Abe2002,
CGY2002, WJ1,WJ2,LB2007, LB2013}. In the pervious works, the fusion
probability was obtained by the passing probability of a Brownian
particle over the top of an inverse harmonic potential
\cite{hof1985,Abe2000}, the latter has been generalized to include
effects of  quantum fluctuation \cite{JDB2002}, initial distribution
\cite{DB2003}, anomalous diffusion \cite{JDB2003} as well as colored
noise \cite{JC2011}.

As one knows that the fusion probability has been estimated by the
stationary value of time-dependent passing probability in terms of
the fusion by diffusion model \cite{Abe2000}, it has a simple form
of error function. There are no need for considering the shell
correction of potential energy and neutron emission in the fusion
phase. Actually, the transient process is very important for the
asymptotical passing probability regarding as the fusion
probability. The inverse harmonic potential approximation is
suitable only for the near barrier fusion and high fission barrier
cases. In this case, the fission life or the mean first passage time
from the ground state to the barrier is much longer than the
transient time of passing probability over the saddle point;
however, the super-heavy element cases should be much carefully,
because the component inside the barrier of time-dependent spatial
distribution function (SDF) decays quickly and opposes to the
process of passing-over barrier. Therefore, it is necessary to
consider the influence of metastable potential structure upon the
passing probability over the barrier. Of course, competition between
neuron emission and fission decay needs to be investigated, the
former decreases the temperature of compound nucleus, but which
occurs in the survival-evaporation phase. At present, we focus on
time-dependent dynamical fusion probability modified by the effect
of reflection boundary of metastable potential.

The paper is organized as follows. In Sec. \ref{sec:level2}, we
describe the barrier passage dynamics and propose an approximate
expression for the probability passing and against escaping over the
barrier of metastable potential. In this section, we also analyze
the error for the stationary passing probability over the saddle
point of inverse harmonic potential regarded as the fusion
probability. Finally, concluding remarks are given in Sec.
\ref{sec:level3}.

\begin{figure}
\includegraphics[scale=0.74]{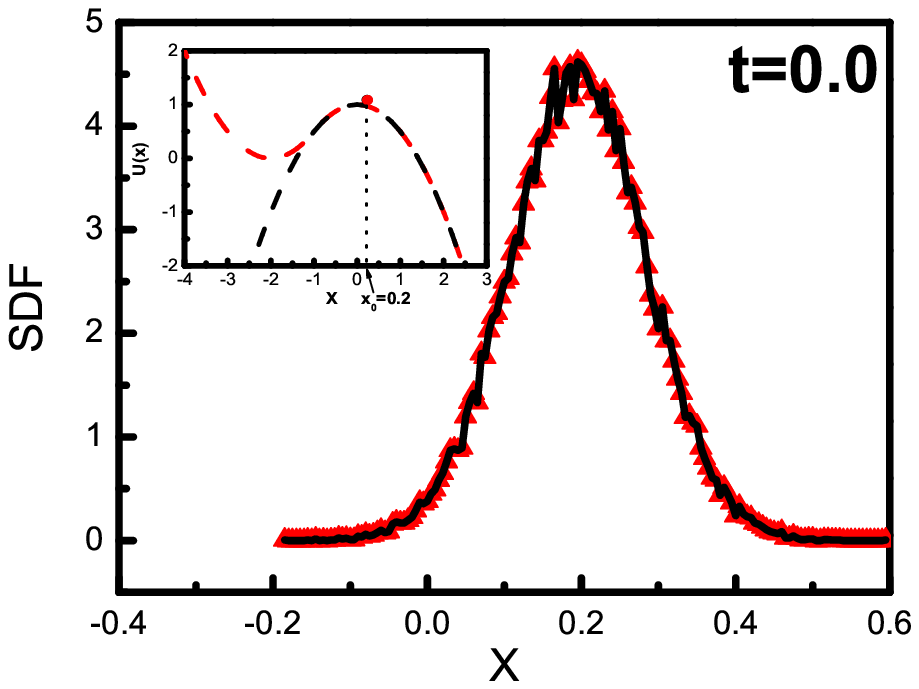}%\label{Fig2}
\includegraphics[scale=0.74]{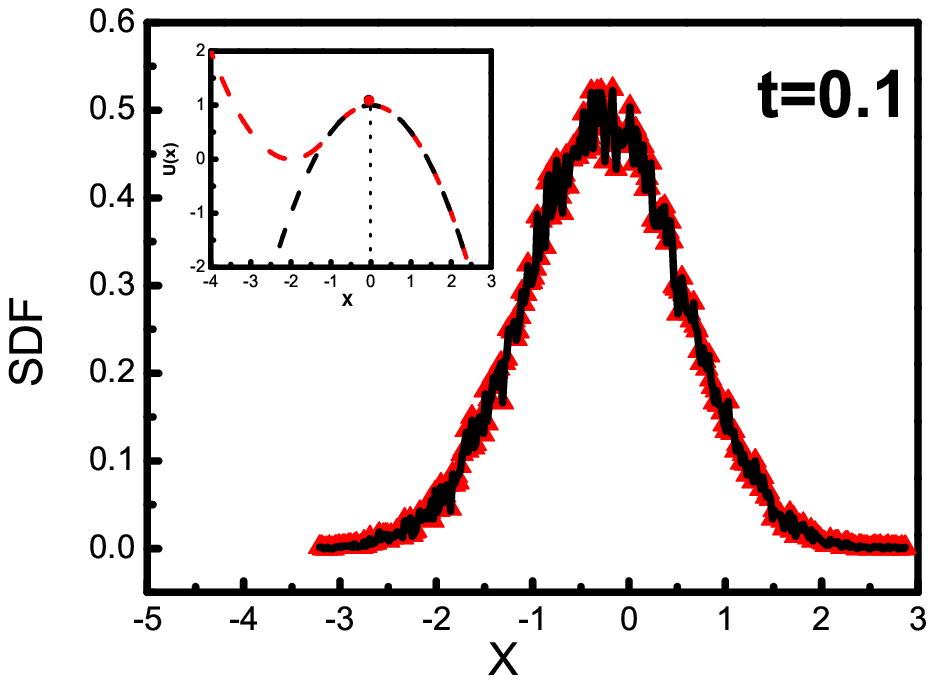}
\includegraphics[scale=0.74]{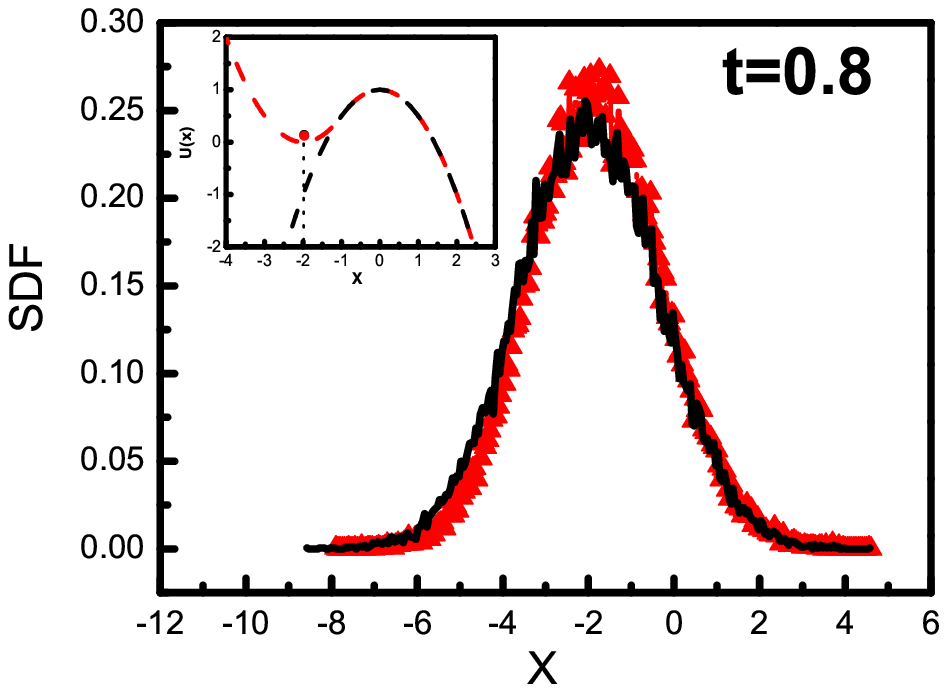}%\label{Fig2}
\includegraphics[scale=0.74]{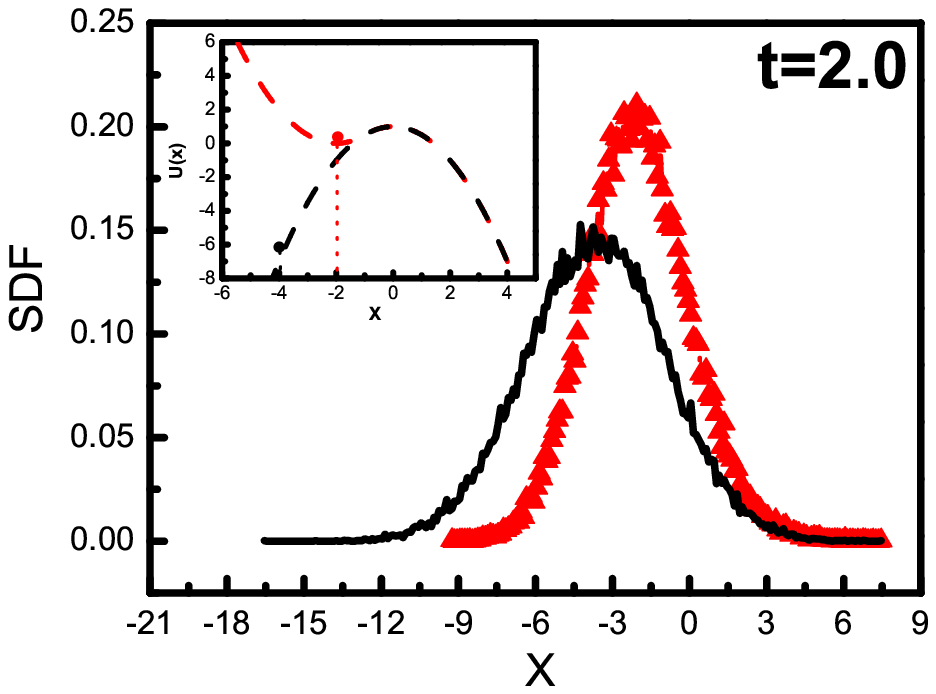}
\includegraphics[scale=0.74]{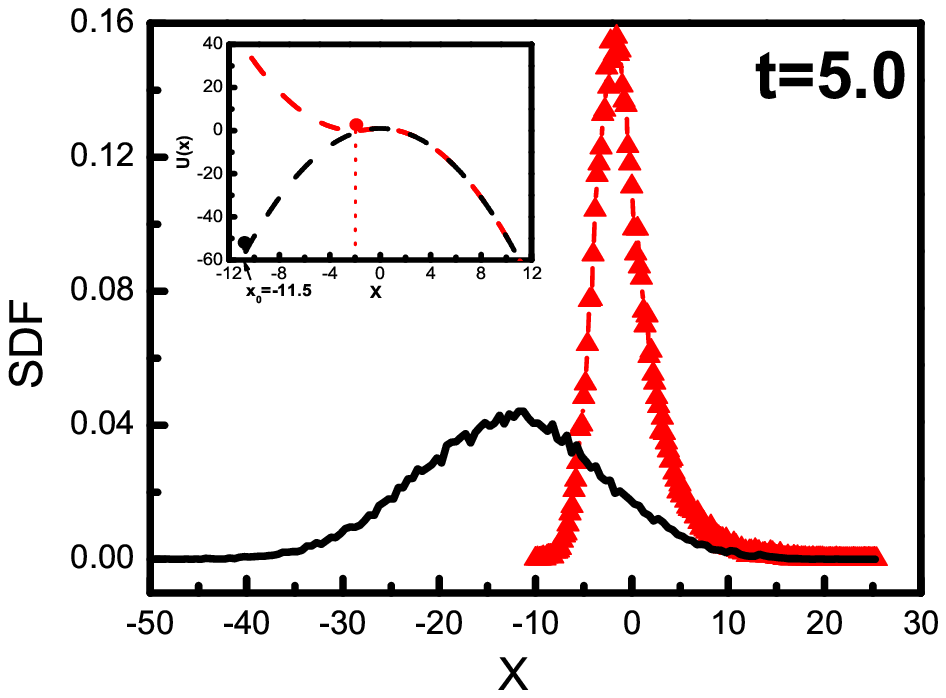}%\label{Fig2}
\includegraphics[scale=0.74]{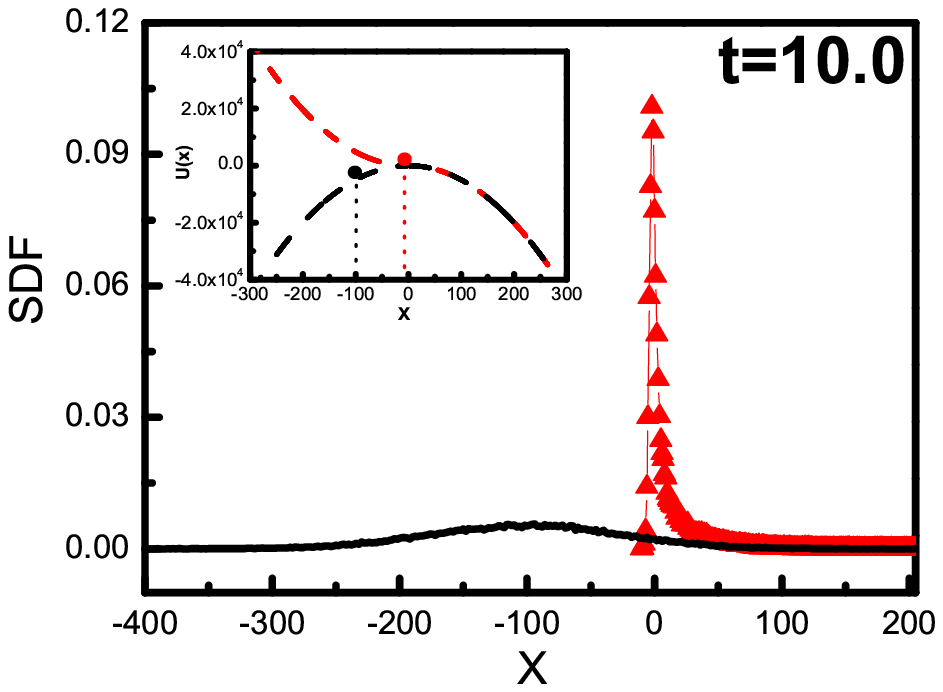}
\caption{\label{Fig1}(color online). Time evolution of SDF of a
particle. The black-solid and red-triangle lines are the SDFs of
particle in the inverse harmonic and metastable potentials,
respectively. Each inset shows the potential with dots representing
the positions where the peak of the distributions locate. The
parameters used are: $T=0.4$, $\gamma=1.0$, $U_b=1.0$,
$\bar{v}_0=-5.0$ and $\bar{x}_0=0.2$. Note that each subgraph has a
different scale.}
\end{figure}

\section{\label{sec:level2}the modified barrier passing probability and fusion-fission dynamics}

The dynamics of a Brownian particle of mass $m$ subjected to a
fluctuation force $\xi(t)$ in a potential $U(x)$ is described by
the following Langevin equation:
\begin{equation}
m\ddot{x}(t)+\gamma\dot{x}(t)+U'(x)=\xi(t), \label{LE}
\end{equation}
where $\xi(t)$ is the Gaussian white noise satisfying
$\langle\xi(t)\rangle=0$ and
$\langle\xi(t)\xi(t')\rangle=2\gamma{k_B}{T}\delta(t-t')$, $k_B$ is
the Boltzmann constant, $T$ is the temperature and $\gamma$ is the
damping coefficient. In order to present an approximate expression
for time-dependent passing probability and against escaping over the
barrier in a metastable potential, we consider an inverse harmonic
potential linking smoothly with a harmonic potential,
\begin{equation}
U(x)=\left\{
\begin{array}{ll}
  U_g(x)=\frac{1}{2}\omega_g^2(x-x_g)^2,&\mbox{$x\le x_c$};\\
  U_s(x)=U_b-\frac{1}{2}\omega_s^2x^2,&\mbox{$x\ge x_c$}.
\end{array}
\right. \label{potential}
\end{equation}
where $x_g$ denotes the coordinate of the ground state, $\omega_g$
and $\omega_s$ are the circular frequencies of potential at the
ground state and the saddle point, respectively, the linking point
of two potentials is determined by
$x_c=x_g\omega_g^2/(\omega_g^2+\omega_s^2)$ through
$U_g(x_c)=U_s(x_c)$ and $U'_g(x_c)=U'_s(x_c)$, $U_b$ is the barrier
height given by $U_b=\frac{1}{2}\omega_s^2x_cx_g$. In the
calculations, all the parameters are chosen to be dimensionless
forms and $m=k_B=1.0$. By the way we choose $x_s=0$ to be the
coordinate of saddle point.

Firstly, in Fig. \ref{Fig1}, we use Langevin Monte-Carlo simulation
to plot time evolution of SDF of the particle in the inverse
harmonic potential and the metastable potential, respectively. It is
seen that the two SDFs are the same at the beginning, because the
metastable well does not bring effect; however, some test particles
have come into the saddle point and then the both occur different,
as time goes. Due to the reflection boundary of metastable
potential, the SDF in the potential of this kind shows
quasi-stationary Boltzmann distribution around the well and its
right-tail escapes continually from the barrier, of course, all the
test particles escape form the barrier in the long time limit.
Nevertheless, the SDF in the inverse harmonic potential case retains
Gaussian all along, but its center tends towards to the infinity
after crossing over the potential top when the initial conditions
are larger than the critical conditions \cite{Abe2000}. On the other
hand, we find that the descent time of the particle from the barrier
to the bottom of well is enough fast, so that the influence of this
precess upon the modified passing probability is not important.

Let us reconsider the time-dependent process for passing over the
saddle point of an inverse harmonic potential, in this case, the
first equation in Eq. (\ref{potential}) is ignored. This model
 has been used widely in
the calculations of fusion probability. The Brownian particle
locals initially at the position $x_0>0$ and has a negative
velocity $v_0<0$. The phase distribution function $W(x,v,t)$ of
the particle at time $t$ is also a Gaussian one due to both linear
equation and Gaussian noise, it is written as \cite{DB2003,
DY2006, YCGDB2008, JDB2003, BY2008}:
\begin{equation}
W(t;x,v)=\frac{1}{2\pi\sigma_x(t)\sigma_{v(t)}}\exp\bigg(-\frac{[x(t)-\langle
x(t)\rangle]^2}{2\sigma^2_x(t)}\bigg)\exp\bigg(-\frac{[v(t)-\langle
v(t)\rangle]^2}{2\sigma^2_v(t)}\bigg), \label{pdf1}
\end{equation}
 where $\langle x(t)\rangle$ is the average position of the
 particle
and $\sigma_x^2(t)$ is the coordinate variance, they are
respectively \cite{DB2003}
\begin{gather}
\langle x(t)\rangle=x_0A(t)+v_0B(t),\\
\sigma_x^2(t)=\frac{T}{m\omega_s^2}\Bigg\{\exp(-\gamma
t)\Bigg[\frac{2\gamma^2}{4\omega_s^2+\gamma^2}
\sinh^2\bigg(\frac{t}{2}\sqrt{4\omega_s^2+\gamma^2}\bigg)\nonumber\\
+\frac{\gamma}{\sqrt{4\omega_s^2+\gamma^2}}\sinh\bigg(t\sqrt{4\omega_s^2+\gamma^2}
\bigg)+1\Bigg]-1\Bigg\},
\end{gather}
where $A(t)$ and $B(t)$ are given by
\begin{eqnarray}
A(t)&=&\exp(-\gamma
t)\Bigg[\cosh\bigg(\frac{t}{2}\sqrt{4\omega_s^2+\gamma^2}\bigg)+\frac{\gamma}{\sqrt{4\omega_s^2+\gamma^2}}
\sinh\bigg(\frac{t}{2}\sqrt{4\omega_s^2+\gamma^2}\bigg)\Bigg], \nonumber\\
B(t)&=&\frac{2}{m\sqrt{4\omega_s^2+\gamma^2}}\exp(-\gamma
t)\sinh\bigg(\frac{t}{2}\sqrt{4\omega_s^2+\gamma^2}\bigg).
\end{eqnarray}

Time-dependent passing probability $P_{\textmd{pass}}(t,x_0,v_0)$
of the particle over the saddle point of inverse harmonic
potential is determined by
\begin{eqnarray}
P_{\textmd{pass}}(t;x_0,v_0)=\int_{-\infty}^{\infty}\int_{-\infty}^{0}W(t;x,v)dvdx
=\frac{1}{2}{\rm erfc}\bigg(\frac{\langle
x(t)\rangle}{\sqrt{2}\sigma_x(t)}\bigg)\label{inverse},
\end{eqnarray}
which depends on the initial preparations of coordinate and velocity
of the particle.

In the case of heavy-ions fusion, a dispersion of the initial
conditions should be considered with a different width, assuming a
Gaussian distribution \cite{DB2003},
\begin{equation}
W_0(\bar x_0, \sigma_{x_0}, \bar v_0,
T_0)=\frac{1}{2\pi\sigma_{x_0}\sqrt{mT_0}}\exp\bigg(-\frac{[x_0-\bar
x_0]^2}{2\sigma_{x_0}^2}\bigg)\exp\bigg(-\frac{[v_0-\bar
v_0]^2}{2mT_0}\bigg). \label{pdf2}
\end{equation}
Thus time-dependent passing probability $\bar
P_{\textmd{pass}}(t,x_0,v_0)$ over the saddle point of inverse
harmonic potential is written as
\begin{eqnarray}
\bar P_{\textmd{pass}}(t;\bar x_0, \sigma_{x_0}, \bar v_o,
T_0)&=&\int_{-\infty}^{\infty}dx_0\int_{-\infty}^{\infty}dv_0P_{\textmd{pass}}(t;x_0,v_0)W_0(\bar
x_0, \sigma_{x_0}, \bar v_o, T_0)\nonumber\\
    &=& \frac{1}{2}{\rm
erfc}\bigg(\frac{\langle \bar
x(t)\rangle}{\sqrt{2}\sigma_x'(t)}\bigg)\label{inverse1},
\end{eqnarray}
where $\langle \bar x(t)\rangle$ is the same as in Eq. (5), provided
that $x_0$ and $v_0$ are replaced by $\bar x_0$ and $\bar v_0$,
respectively. The variance becomes
\begin{equation}
\sigma_x'^2(t)=\sigma_x^2(t)+\sigma_{x_0}^2(t)A^2(t)+mT_0B^2(t).
\end{equation}
In these equations, $T_0$ is a parameter for the initial
distribution that could be interpreted as the temperature of the
nuclei at contact \cite{DB2003}. Naturally, the SDF of particle
under fluctuation force becomes wider and wider, its center moves
along the direction of initial velocity, as time goes. After the
transient time, a part of the SDF has passed over the saddle point
and then the passing probability converges to a finite value with
$0\leq \bar P_{\textmd{pass}}\leq 1$, because of
$\lim_{t\to\infty}\langle \bar x(t)\rangle/\sigma_x'(t)$=constant
in Eq. (10).

\begin{figure}
\includegraphics[scale=0.70]{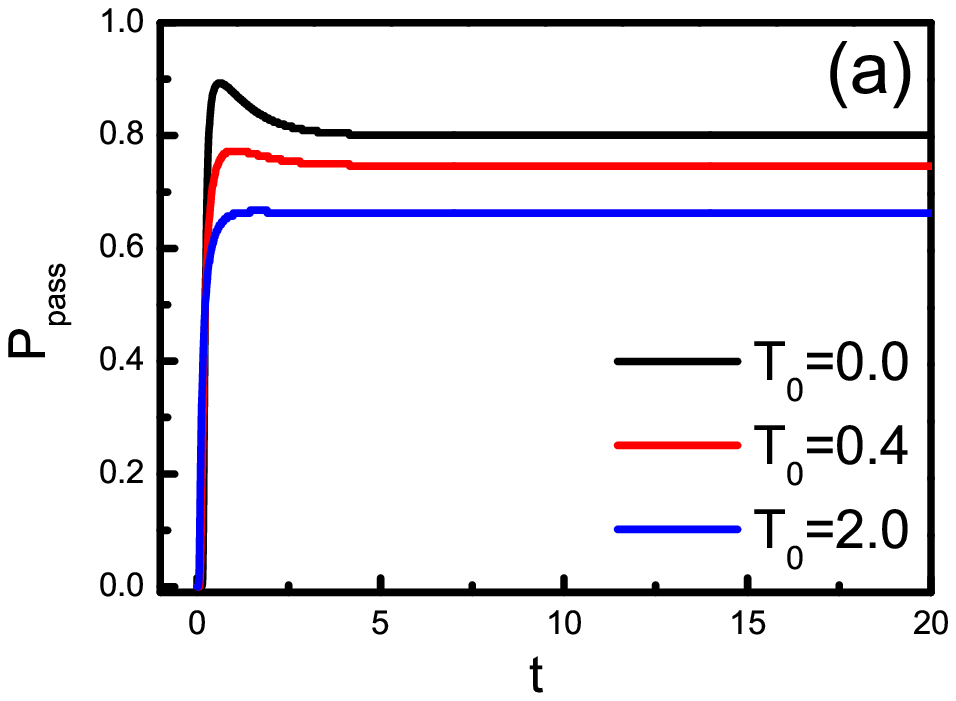}
\includegraphics[scale=0.70]{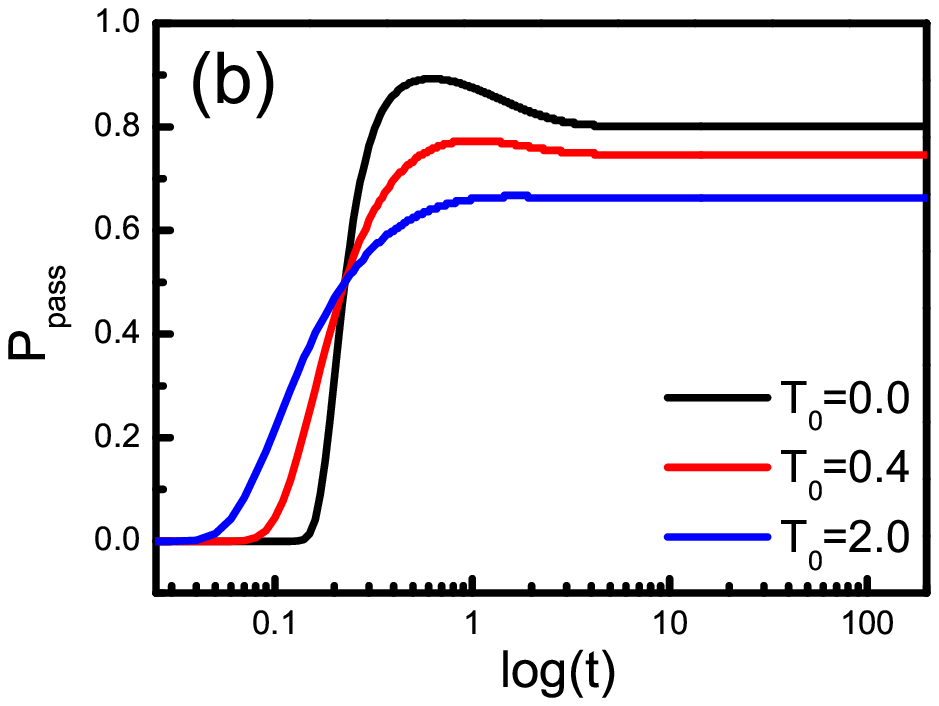}
\caption{\label{Fig2}(color online). (a) Time-dependent passing
probability over the saddle point of inverse harmonic potential with
three kinds of typical initial temperature of thermalization
($T_0=0.0$, $T_0=0.4$ and $T_0=2.0$) and (b) is the time-dependent
passing probability with logarithmic of the time. Here, $T=0.4$,
$\gamma=1.0$, $U_b=1.0$, $\sigma_{x_0}=0.0$, $\bar{v}_0=-5.0$ and
$\bar{x}_0=0.2$.}
\end{figure}

In Fig. \ref{Fig2}(a), we can see that the stable value of the
time-dependent passing probability decreases with the increase of
the initial temperature of thermalization. We also find that the
descent time of the particle from the barrier to the bottom of well
increases with the increase of the initial temperature of
thermalization, as shown in Fig. \ref{Fig2}(b). As a consequence,
the initial kinetic energy should be considered into account and
this result is similar to the ref. \cite{DB2003} for a sharp initial
condition.

We now address the modified passing probability taking into account
the influence of reflection boundary of potential by a reasonable
assumption. According to the Kramers rate theory, the particle
subjected to thermal fluctuation in the metastale potential will
decay over the barrier finally \cite{MG1993,DY2006}. We multiply the
exponential decay factor into the passing probability which has been
coupled the fusion and fission processes, so that the modified
passing probability, namely, time-dependent probability of the
particle staying inside the saddle point, is assumed to be
\begin{eqnarray}
P_{\textmd{m-pass}}(t;x_0,v_0)=\bar P_{\textmd{pass}}(t;\bar x_0,
\sigma_{x_0}, \bar v_o, T_0)\exp(-r_et)
            = \frac{1}{2}{\rm
erfc}\bigg(\frac{\langle \bar
x(t)\rangle}{\sqrt{2}\sigma_x'(t)}\bigg)\exp(-r_et),\label{survival}
\end{eqnarray}
where $r_e$ is the steady escape rate
\cite{HA1940,PH1990,PH1988,RJ1980,VS2006,DABK2007,PT1994}. This
approximation implies that once the particle passes over the barrier
top at last time, it should escape over the barrier with the Karmers
decay form.
  In Fig. \ref{Fig2}(b), it is obviously that the transient time can be
ignored in the calculation of the time-dependent modified passing
probability.
 If $r_e\to 0$, $\exp(-r_et)\simeq 1$ after a finite
time, the modified passing probability [Eq. (\ref{survival})]
approaches the passing probability [Eq. (\ref{inverse1})] for the
inverse harmonic potential.

\begin{figure}
\includegraphics[scale=0.70]{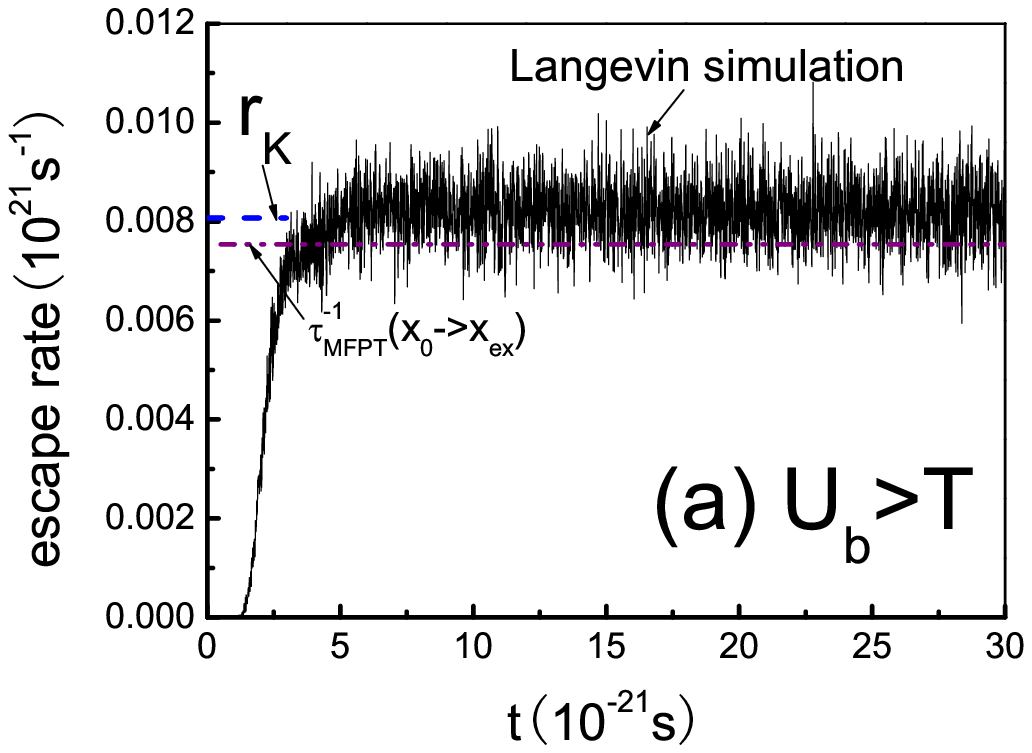}
\includegraphics[scale=0.70]{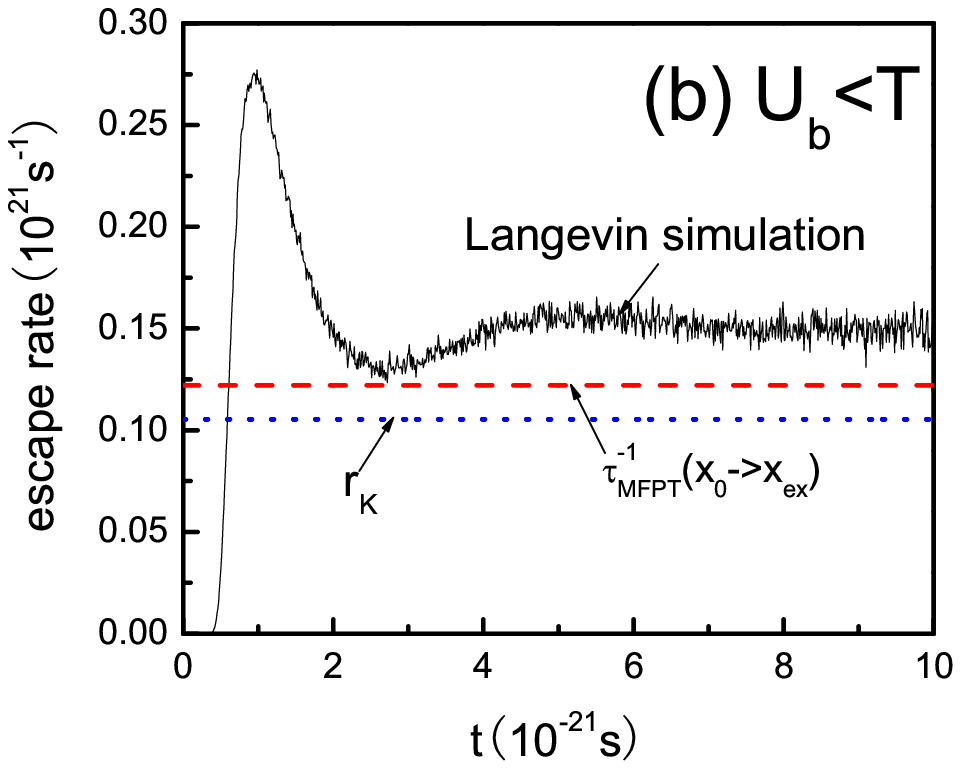}
\caption{\label{Fig3}(color online). Time-dependent escape rate
calculated by Langevin simulation and compared by the analytical
formula of two kinds. (a) is the low-temperature case ($U_b=1.0$,
$T=0.4$) and (b) is the high-temperature case ($U_b=0.25$,
$T=2.0$).}
\end{figure}

The Kramers rate formula \cite{HA1940,PH1990,VS2006} produces the
better stationary result of time-dependent escape rate when the
barrier height of metastable potential is larger than the
temperature, as shown in Fig. \ref{Fig3}(a). However, when the
temperature is larger than the barrier height, the Kramers rate
formula is not applicable, we use the inverse of the mean first
passage time (MFPT) \cite{DABK2007} across an exit $x_{\textmd{ex}}$
given by
\begin{equation}
\tau_{\textmd{MFPT}}(x_0\to
x_{\textmd{ex}})=\Bigg(\frac{\sqrt{\frac{\gamma^2}{4}+\omega_s^2}-\frac{\gamma}{2}}{\omega_s}\Bigg)^{-1}
\frac{\omega_s}{T}\int_{x_0}^{x_{\textmd{ex}}}dy\exp\bigg[\frac{U(y)}{T}\bigg]
\int_{-\infty}^{y}dz\exp\bigg[-\frac{U(z)}{T}\bigg],\label{MFPT}
\end{equation}
 to replace of the stationary escape rate of
particle in a metastable potential well, i.e.,
$r_e=(\tau_{\textmd{MFPT}})^{-1}$ \cite{PH1990,PT1994}. Noticed that
we introduce here a correction factor of general damping to the
pervious overdamped result, indeed, Eq. (\ref{MFPT}) is in agreement
with the result of Refs.
\cite{JY2004,SKK1958,HF2003,HA2003,DBC2004,DABK2007} in the
overdamped case ($\gamma\gg\omega_s$).  At low temperature, Eq.
(\ref{MFPT}) can be evaluated within the steepest-descent
approximation \cite{PH1990} as the following
\begin{equation}
\tau_{\textmd{MFPT}}(x_0\to
x_{\textmd{ex}})=\Bigg(\frac{\sqrt{\frac{\gamma^2}{4}+\omega_s^2}-\frac{\gamma}{2}}
{\omega_s}\Bigg)^{-1}\frac{2\pi}{\omega_g}\exp\bigg(\frac{U_b}{T}\bigg),
\end{equation}
its inverse coincides with the Kramers rate formula
\cite{CWG1985}.

\begin{figure}
\includegraphics[scale=0.80]{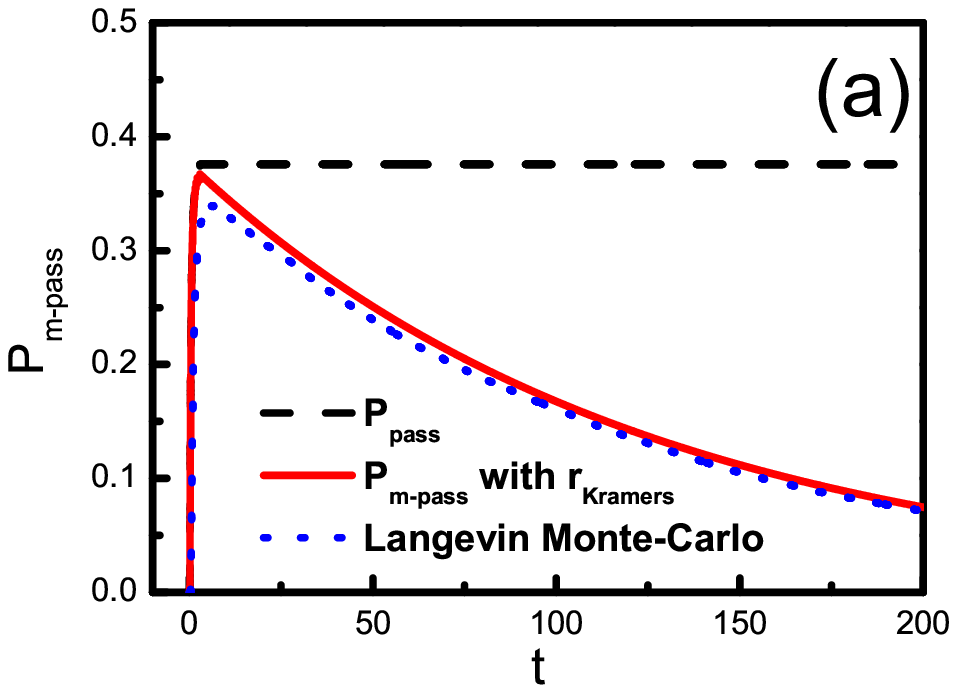}\includegraphics[scale=0.80]{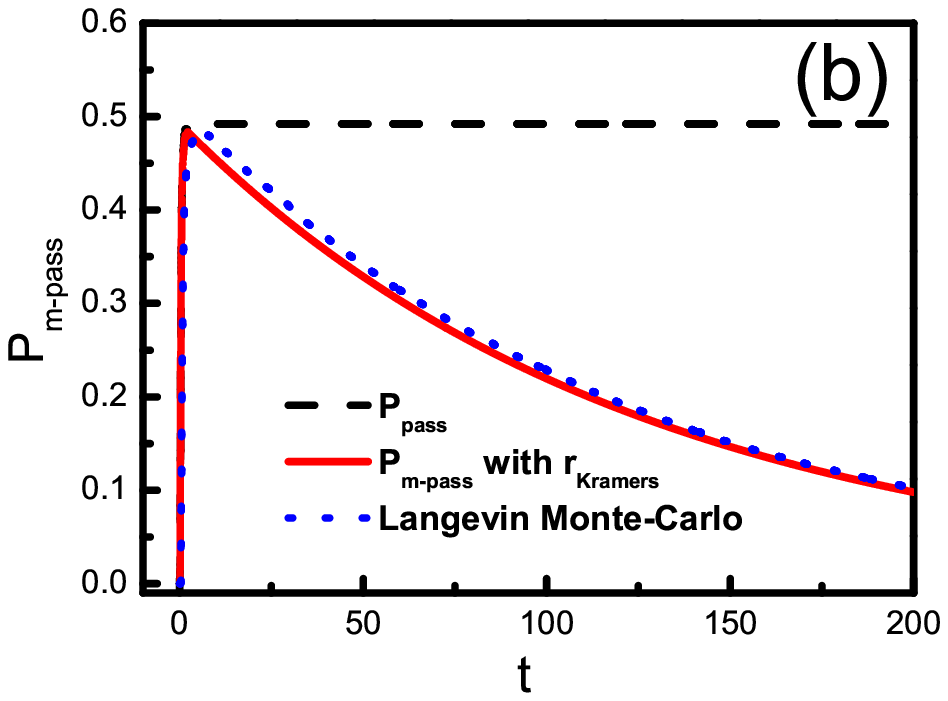}
\includegraphics[scale=0.80]{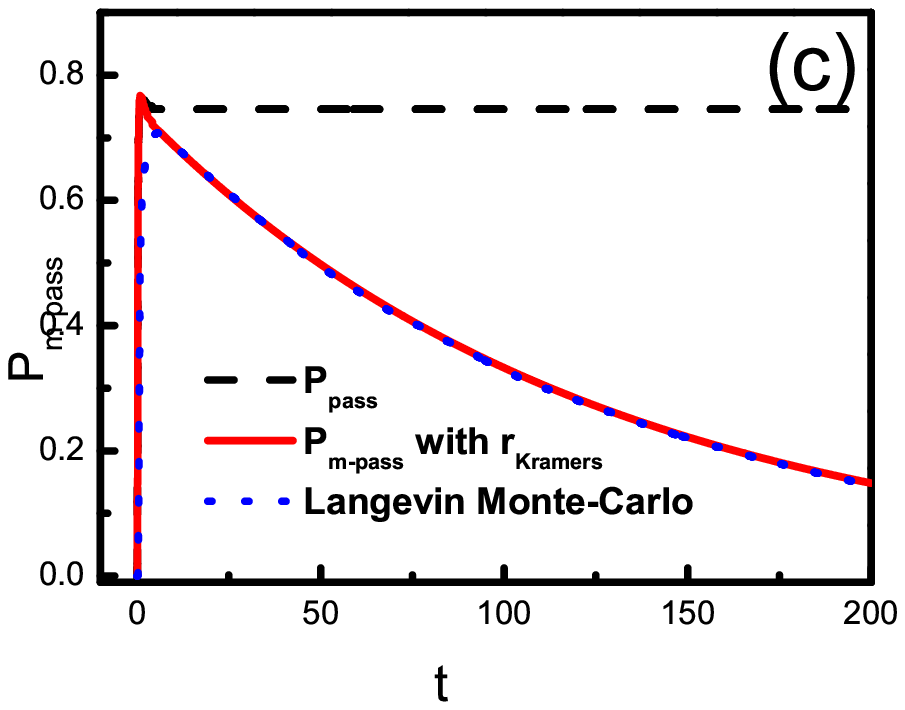}
\caption{\label{Fig4}(color online). The time-dependent modified
passing probability over the barrier of metastable potential and the
passing probability over the saddle point of inverse harmonic
potential. The parameters used are: $U_b=1.0$, $T=0.4$, $T_0=0.4$,
$\sigma_{x_0}=0$, $\gamma=1.0$, $x_0=0.2$.}
\end{figure}

Furthermore, the advance of MFPT or the mean last passage time
(MLPT) \cite{JY2004} is not restricted to smooth metastable
potentials, Eq. (\ref{MFPT}) is still suitably even if the nuclear
shell correction is taken into account in the deformation potential
energy of super-heavy elements. A statistical proof for the relation
between the Karmers rate constant and the MFPT or the MLPT was
presented in Ref. \cite{DBC2004}.

\begin{figure}
\includegraphics[scale=0.80]{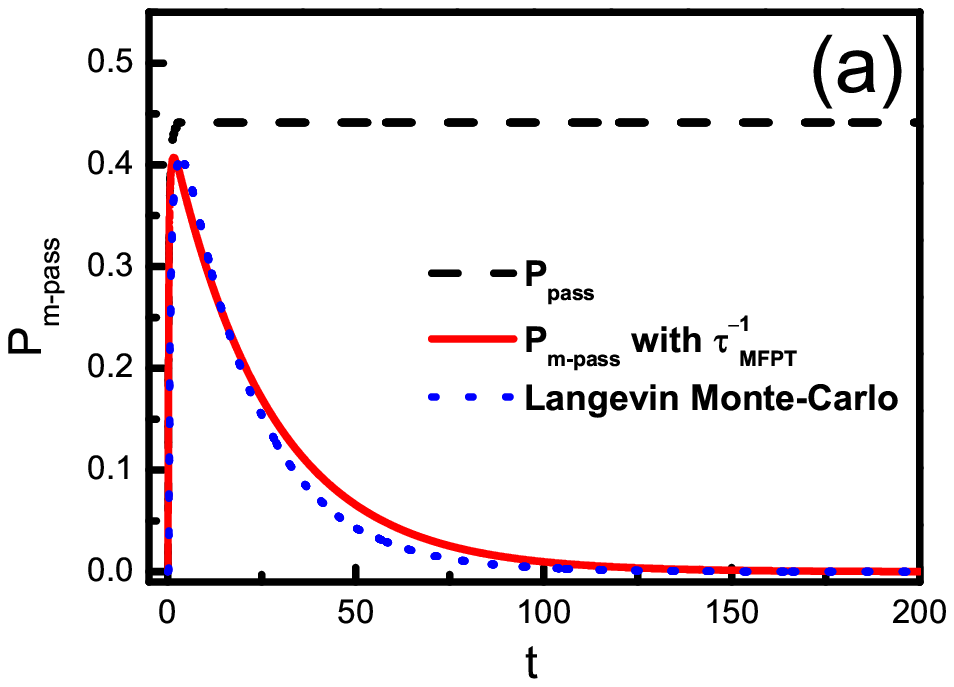}\includegraphics[scale=0.80]{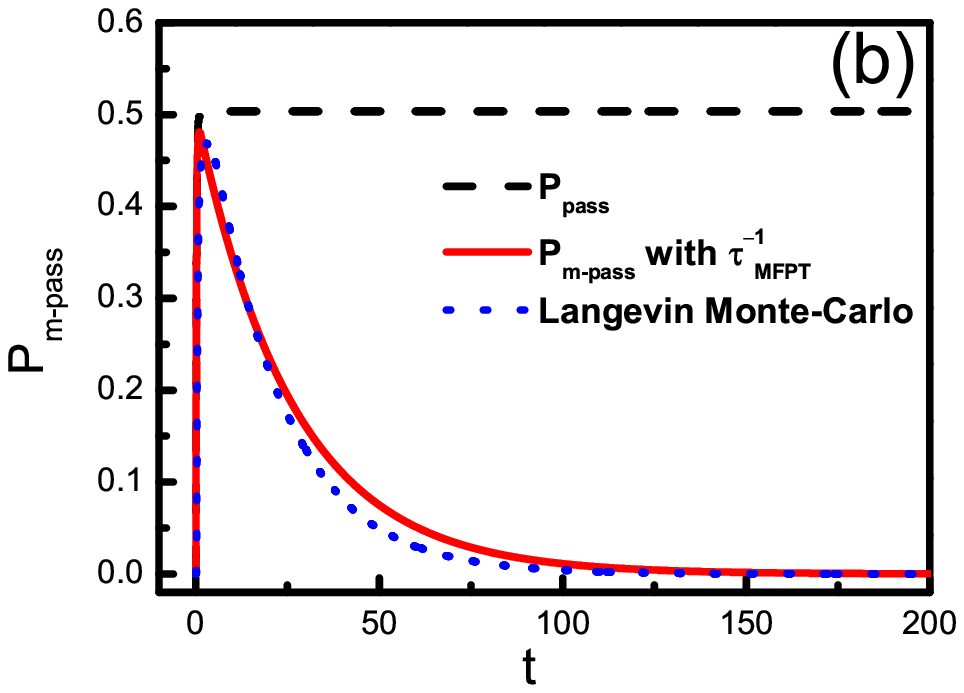}
\includegraphics[scale=0.80]{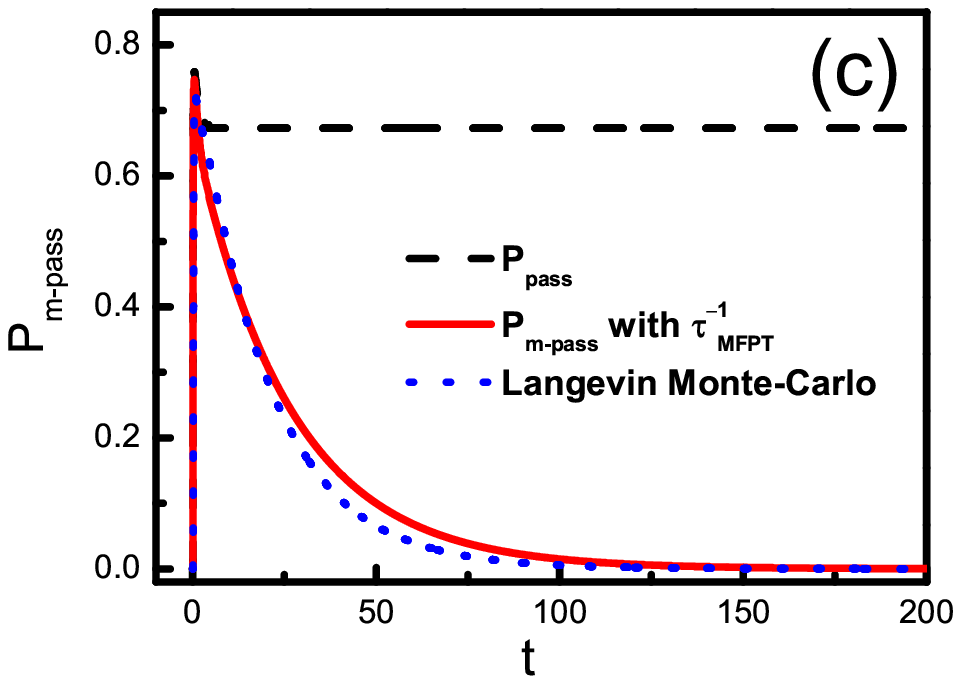}
\caption{\label{Fig5}(color online). Comparison of time-dependent
modified passing probability over the barrier of metastable
potential and the passing probability over the saddle point of
inverse harmonic potential. The parameters used are: $U_b=1.0$,
$T=2.0$, $T_0=2.0$, $\gamma=2.0$, $\sigma_{x_0}=0$, $x_0=0.2$.}
\end{figure}

In Figs. \ref{Fig4} and \ref{Fig5}, we compare the time-dependent
modified passing probability [Eq. (\ref{survival})] with the
Langevin Monte-Carlo simulation for Eqs. (\ref{LE}) and
(\ref{potential}) and the passing probability [Eq. (\ref{inverse1})]
over the saddle point of inverse harmonic potential, respectively,
where three typical initial velocities are used. It is evident from
Eq. (\ref{survival}), the modified passing probability over the
barrier of metastable potential approaches zero in the long-time
limit.

It is seen from Fig. \ref{Fig4} that the modified passing
probability calculated by our theoretical formula is in agreement
with the Langevin Monte-Carlo simulation when $U_b>T$. In
particular, the maximal value of time-dependent modified passing
probability is close to the stationary value of passing probability
over the saddle point of inverse harmonic potential. This means that
the influence of reflection boundary of metastable potential upon
the transient part of time-dependent passing probability is weakly
in the case of low temperature or high barrier. Figure \ref{Fig5}
shows the calculated result at high temperature, in which the
barrier height of metastable potential is $U_b=1.0$ and the
temperature $T=2.0$.

If the barrier height is low, the particle under influence of
reflection boundary of potential is easier to escape over the
saddle point, so that the time required for the modified passing
probability arriving at the maximum is earlier than that of the
passing probability approaching its stationary value. This
concludes that the reflection boundary of metastable potential
plays a decreasing role to the transient result of passing
probability.

We have proposed the expression of time-dependent  modified
passing probability against escaping over the barrier of the
metastable potential, i.e., Eq. (\ref{survival}), the time leading
to $P_{\textmd{m-pass}}$ become the maximum is determined by the
positive real root of following equation:
\begin{eqnarray}
\frac{dP_{\textmd{m-pass}}}{dt}=\frac{1}{2}\exp(-r_et)J(t)-\frac{1}{2}r_e
                   \exp(-r_e t){\rm erfc}\bigg(\frac{\langle
                   \bar
                   x(t)\rangle}{\sqrt{2}\sigma_x'(t)}\bigg)=0,\label{derive}
\end{eqnarray}
where $J(t)$ is the derivative of ${\rm erfc}[\langle \bar
x(t)\rangle/(\sqrt{2}\sigma_x'(t))]$ given by
\begin{equation}
J(t)=-\frac{2}{\sqrt{\pi}}\exp\bigg(-\frac{\langle \bar
x(t)\rangle^2}{\sigma_x'^2(t)}\bigg)
\bigg[\frac{M(t)}{\sqrt{2}\sigma_x'(t)}-\frac{T}{2\sqrt{2}m\omega_s^2}\frac{\langle
\bar x(t)\rangle G(t)}{\sigma_x'^3(t)}\bigg],
\end{equation}
where $M(t)$ and $G(t)$ are
\begin{eqnarray}
M(t)&=&\exp(-\gamma t)\bigg[\bigg(\frac{\bar v_0}{m}-\frac{\bar x_0
\gamma}{2}\bigg)\cosh(\frac{1}{2} a t)+\bigg(\frac{a \bar
x_0}{2}+\frac{\gamma^2 \bar x_0}{a}-\frac{2\gamma
\bar v_0}{m a}\bigg)\sinh(\frac{1}{2} a t)\bigg], \nonumber\\
G(t)&=&2\gamma \bigg(1-\frac{\gamma^2}{a^2}\bigg)\exp(-\gamma
t)\sinh^2(\frac{1}{2}at),\label{MG}
\end{eqnarray}
where  $a=\sqrt{4\omega_s^2+\gamma^2}$. Hence the maximum of the
time-dependent modified passing probability can be obtained by Eq.
(\ref{survival}) through solving numerically Eqs.
(\ref{derive})-(\ref{MG}). Noticed that this quantity is a defined
one depending on the model parameters. It is seen from Fig.
\ref{Fig2} that the time corresponding to the maximal staying
probability is equal approximately to the transient time of the
passing probability only in the case of high barrier.

\begin{figure}
\includegraphics[scale=0.60]{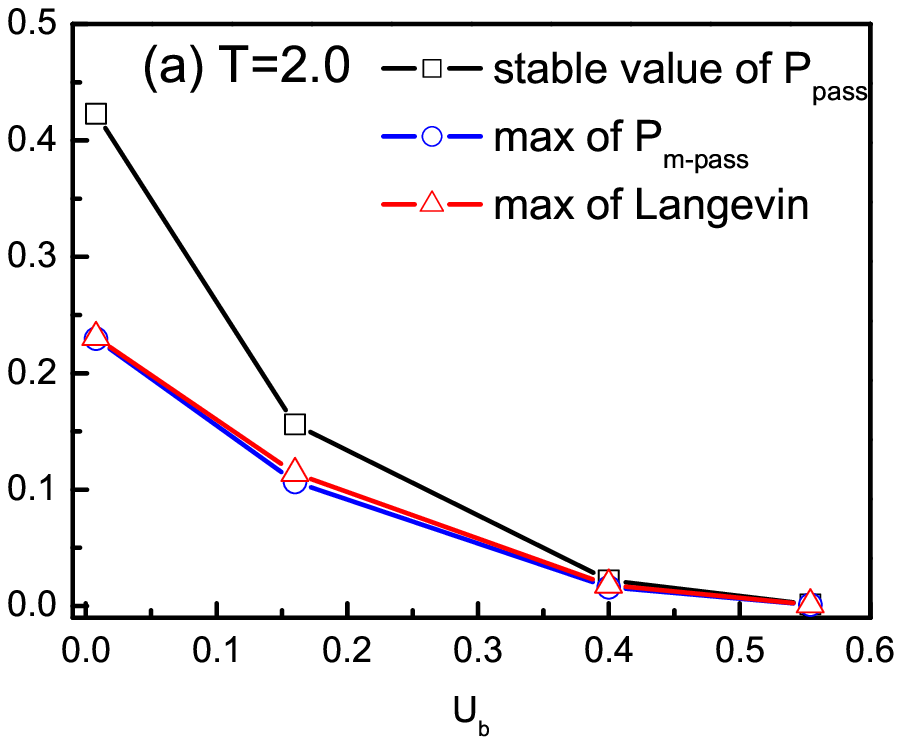}\includegraphics[scale=0.60]{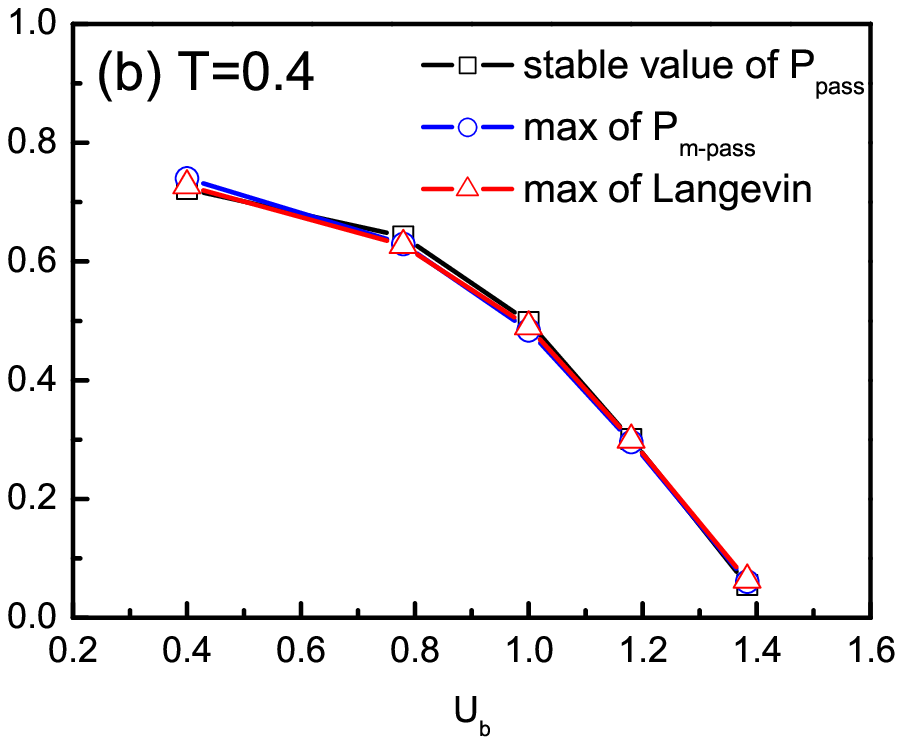}
\caption{\label{Fig6}(color online). The maximum value of
time-dependent modified passing probability (blue-circled-line) over
the barrier of the metastable potential and the stationary passing
probability (black-squared-line) over the saddle point of inverse
harmonic potential, they are compared with the Langevin Monte-Carlo
simulations (red-triangled-line).}
\end{figure}

In Fig.\ref{Fig6}, we show the maximal value of time-dependent
modified passing probability over the saddle point of the metastable
potential as a function of the barrier height, which is also
compared with the stationary passing probability over the saddle
point of inverse harmonic potential.  It is seen that with the
increase of the barrier height, the maximal value of time-dependent
modified passing probability is close to the stationary value of the
passing probability over the saddle point of inverse harmonic
potential, so that one can approximately treat the asymptotical
passing probability over the saddle point of the inverse harmonic
potential as the fusion probability in massive nuclear fusion
reaction. However, when the fission barrier is low, this occurs in
the super-heavy element cases, the stationary value of
time-dependent passing probability over the saddle point of inverse
harmonic potential is no longer applicable for the fusion
probability. From the present work, we think that it is better to
regard the maximal value of time-dependent modified passing
probability over the saddle point of the metastable potential as the
fusion probability, for that the modified passing probability is the
result of whole fusion-fission process.

\section{\label{sec:level3}CONCLUSION}

We have investigated whole fusion-fission process with Langevin
approach, in which the influence of reflection boundary of the
metastable potential is taken into account in the calculation of
time-dependent passing probability over the saddle point. By the
passing probability over the saddle point of inverse harmonic
potential multiplying the exponential decay factor of the particle
in the metastable potential, an approximate analytical expression of
the modified time-dependent passing probability over the saddle
point of metastable potential has been proposed. Our results have
shown that only when the fission barrier of of fusing system is
larger than the temperature, the stationary passing probability over
the saddle point of an inverse harmonic potential can be regarded as
the fusion probability of massive nuclei. Nevertheless, at low
fission barrier, the reflection boundary plays a decreasing role for
the passing probability over the saddle point. It has been found
that the time required for the modified time-dependent passing
probability arriving at the maximal value is earlier than the
transient time of the passing probability. This is due to the
decaying probability against the passing probability.

\section*{ACKNOWLEDGMENTS}

This work was supported by the National Natural Science Foundation
of China under Grant No. 11175021 and the Specialized Research Fund
for the Doctoral Program of Higher Education under Grant No.
20120003110025.


\begin{thebibliography}{References}

\bibitem{HA1940}H. A. Kramers, Physica (Utrecht) \textbf{7}, 284 (1940).


\bibitem{Abe1986}Y. Abe, J. de Physique 47(8), C4-329 (1986).

\bibitem{PH1990}P. H\"anggi, P. Talkner, and M. Borkovec, Rev. Mod. Phys. \textbf{62}, 251
(1990).
\bibitem{YA1997}Y. Aritomo, T. Wada, M. Ohta, and Y. Abe, Phys. Rev. C  \textbf{55}, R1011 (1997).

\bibitem{YY1997}Y. Abe, Y. Aritomo, T. Wada, and M. Ohta, J. Phys. G: Nucl. Part. Phys. \textbf{23}, 1275 (1997).

\bibitem{Abe2002}Y. Abe, D. Boilley, G. Kosenko, J. D. Bao, C. W. Shen, B.
Giraud, and T. Wada, Prog. Theor. Phys. Suppl. \textbf{146}, 104
(2002).

\bibitem{CGY2002}C. W. Shen, G. Kosenko, and Y. Abe, Phys. Rev. C \textbf{66}, 061602(R) (2002).

\bibitem{WJ1} W. J. \'{S}wiatecki, K. Siwek-Wilczy\'{n}ska, and J. Wilczy\'{n}ski,
Acta Phys. Pol. B \textbf{34}, 2049 (2003).

\bibitem{WJ2} W. J. \'{S}wiatecki, K. Siwek-Wilczy\'{n}ska, and J.Wilczy\'{n}ski,
Phys. Rev. C \textbf{71}, 014602 (2005).



%\bibitem{BY2008}B. Yilmaz, S. Ayik, Y. Abe, and D. Boilley, Phys. Rev. E \textbf{77}, 011121 (2008).

\bibitem{LB2006}Z. H. Liu and J. D. Bao, Phys. Rev. C \textbf{74}, 057602 (2006).


\bibitem{LB2007}Z. H. Liu and J. D. Bao, Phys. Rev. C \textbf{76}, 037604 (2007).

\bibitem{LB2013}Z. H. Liu and J. D. Bao, Phys. Rev. C \textbf{87}, 034616 (2013).


\bibitem{hof1985}H. Hofmann and R. Samhammer, Z. Phys. A
\textbf{322}, 157 (1985).

\bibitem{Abe2000}Y. Abe, D. Boilley, B. G. Giraud, and T. Wada, Phys. Rev. E \textbf{61}, 1125 (2000).


\bibitem{JDB2002}J. D. Bao and D. Boilley, Nucl. Phys. A \textbf{707}, 47 (2002).

\bibitem{DB2003}D. Boilley, Y. Abe, and J. D. Bao, Eur. Phys. J. A  \textbf{18}, 627 (2003).

\bibitem{JDB2003}J. D. Bao and Y. Z. Zhuo, Phys. Rev. C  \textbf{67}, 064606 (2003).

\bibitem{JC2011}J. W. Mao and C. W. Shen, Phys. Rev. E \textbf{83}, 041108 (2011).

\bibitem{Liu2011}Z. H. Liu and J. D. Bao, Phys. Rev. C \textbf{83}, 044613 (2011).


\bibitem{Cap2011} T. Cap, K. Siwek-Wilczy\'{n}ska, and J. Wilczy\'{n}ski, Phys.
Rev. C \textbf{83}, 054602 (2011).


\bibitem{YCGDB2008}Y. Abe, C. W. Shen, G. Kosenko, D. Boilley, and B. G. Giraud, Int. J. Mod. Phys. E \textbf{17}, 2214 (2008)

\bibitem{YDBGC2004}Y. Abe, D. Boilley, B. Giraud, G. Kosenko, and C. W. Shen, Acta. Phys. Hung. A \textbf{19}, 77 (2004).

\bibitem{JRMDICK2004}J. O. Newton, R. D. Butt, M. Dasgupta, D. J. Hinde, I. I. Gontchar,
 C. R. Morton, and K. Hagino, Phys. Rev. C \textbf{70}, 024605 (2004).

\bibitem{MI2013}M. V. Chushnyakova and I. I. Gontchar, Phys. Rev. C \textbf{87}, 014614 (2013).

\bibitem{DY2006}D. Boilley and Y. Lallouet, J. Stat. Phys. \textbf{125}, 2 (2006).

\bibitem{BY2008}B. Yilmaz, S. Ayik, Y. Abe, and D. Boilley, Phys. Rev. E \textbf{77}, 011121 (2008).

\bibitem{MG1993}M. Gitterman and G. H. Weiss, J. Stat. Phys. \textbf{70}, 107 (1993).

\bibitem{PH1988}P. Talkner and H. Braun, J. Chem. Phys. \textbf{88}, 7537 (1988).

\bibitem{RJ1980}R. F. Grote and J. T. Hynes, J. Chem. Phys. \textbf{73}, 2715 (1980).

\bibitem{VS2006}V. I. Mel'nikov and S. V. Meshkov, J. Chem. Phys. \textbf{85}, 1018 (2006).


\bibitem{DABK2007}D. Boilley, A. Marchix, B. Jurado, and K. H. Schmidt, Eur. Phys. J. A \textbf{33}, 47 (2007).

\bibitem{PT1994}P. Talkner, Chem. Phys. \textbf{180}, 199 (1994).

\bibitem{SKK1958}S. K. Kim, J. Chem. Phys. \textbf{28}, 1057 (1985).

\bibitem{HF2003}H. Hofmann and F. A. Ivanyuk, Phys. Rev. Lett. \textbf{90}, 132701 (2003).

\bibitem{HA2003}H. Hofmann and A. G. Magner, Phys. Rev. C \textbf{68}, 014606 (2003).

\bibitem{JY2004}J. D. Bao and Y. Jia, Phys. Rev. C \textbf{69}, 027602 (2004).


\bibitem{DBC2004}D. Boilley, B. Jurado, and C. Schmitt, Phys. Rev. E \textbf{70}, 056129 (2004).

\bibitem{CWG1985}H. Risken, \emph{The Fokker-Planck Equation} (Springer, Berlin,
1989); C. W. Gardiner, \emph{ Handbook of Stochastic Methods}
(Springer, Berlin, 2002).





\end{thebibliography}
\end{document}